\documentclass{aa}
\usepackage[varg]{txfonts}

\usepackage{natbib}
\bibpunct{(}{)}{;}{a}{}{,} 

\begin{document}

\title{Evaluating the gravitational wave detectability of globular clusters and the Magellanic Clouds for LISA}

\author{Wouter G. J. van Zeist\inst{1,2,3}\thanks{e-mail: wouter.vanzeist@astro.ru.nl}
\and Gijs Nelemans\inst{1,4,5}
\and Simon F. Portegies Zwart\inst{2}
\and Jan J. Eldridge\inst{3}}

\institute{Department of Astrophysics/IMAPP, Radboud University, PO Box 9010, 6500 GL, Nijmegen, The Netherlands
\and Leiden Observatory, Leiden University, PO Box 9513, 2300 RA, Leiden, The Netherlands
\and Department of Physics, University of Auckland, Private Bag 92019, Auckland, New Zealand
\and SRON, Netherlands Institute for Space Research, Niels Bohrweg 4, 2333 CA, Leiden, The Netherlands
\and Institute of Astronomy, KU Leuven, Celestijnenlaan 200D, B-3001, Leuven, Belgium}

\date{Received XXX / Accepted YYY}

\abstract{We use the stellar evolution code \textsc{bpass} and the gravitational wave simulation code \textsc{legwork} to simulate populations of compact binaries that may be detected by the in-development space-based gravitational wave (GW) detector LISA. Specifically, we simulate the Magellanic Clouds and binary populations mimicking several globular clusters, neglecting dynamical effects. We find that the Magellanic Clouds would have a handful of detectable sources each, but for globular clusters the amount of detectable sources would be less than one. We compare our results to earlier research and find that our predicted numbers are several tens of times lower than calculations using the stellar evolution code \textsc{bse} that take dynamical effects into account, but also calculations using the stellar evolution code \textsc{SeBa} for the Magellanic Clouds.  This correlates with earlier research which compared \textsc{bpass} models for GW sources in the Galactic disk with \textsc{bse} models and found a similarly sized discrepancy. We analyse and explain this discrepancy as being caused by differences between the stellar evolution codes, particularly in the treatment of mass transfer and common-envelope events in binaries, where in \textsc{bpass} mass transfer is more likely to be stable and tends to lead to less orbital shrinkage in the common-envelope phase than in other codes. This difference results in fewer compact binaries with periods short enough to be detected by LISA existing in the \textsc{bpass} population. For globular clusters, we conclude that the impact of dynamical effects is uncertain from the literature, but the differences in stellar evolution have an effect of a factor of a few tens.}

\keywords{gravitational waves -- globular clusters: general -- Magellanic Clouds -- binaries: close -- white dwarfs -- stars: neutron}

\titlerunning{Evaluating the GW detectability of GCs and Magellanic Clouds}
\authorrunning{W. G. J. van Zeist et al.}

\maketitle

\section{Introduction}

\subsection{LISA}

The Laser Interferometer Space Antenna \citep[LISA;][]{lisa_l3,lisa_redbook} is a space-based astronomical observatory that is currently under development, which aims to detect gravitational waves (GWs) in a lower frequency band than those currently being detected by the ground-based LIGO–Virgo–KAGRA (LVK) Collaboration \citep{gwtc3}. It will consist of three satellites in a triangular formation with a separation of 2.5 million km, which will rotate about their barycentre, while the barycentre will follow Earth's orbit but trailing by approximately 20 degrees \citep[see][]{cornish2017}.

There are many different types of GW sources that LISA has been predicted to be able to detect, ranging from those of astrophysical origin like merging binary systems \citep{lisa_astrophysics} to those of cosmological origin like cosmic strings and inflation \citep{lisa_cosmology}. Unlike the LVK Collaboration detectors, which currently have not detected more than one GW signal at a time, it is expected that LISA would continuously receive multiple overlapping signals, and disentangling these will be a complex process \citep{lisa_data_analysis}.

To understand the signals that LISA will receive, it is beneficial to understand and model the different types of sources that LISA might detect, and the populations that exist of these sources. One class of astrophysical sources for LISA will be compact binaries of stellar origin. These are binary systems consisting mostly of two compact objects, which can be either a black hole (BH), neutron star (NS) or white dwarf (WD). It is known that such binaries exist and can be detected in GWs because the GW signals that LVK have observed \citep[catalogued most recently in][]{gwtc3} have been from the mergers of such binaries. LISA, having a lower frequency band, would detect such systems earlier in their evolution, when they are still at greater separations. Further evidence for LISA-detectable binaries is given by the LISA verification binaries \citep[catalogued most recently in][]{kupfer2024}, a set of Galactic binaries known from electromagnetic (EM) observations, predominantly containing WDs, which based upon their frequencies and distances would potentially be detectable by LISA.

There have been many studies about the populations of stellar compact binaries that LISA may detect \citep[e.g.][]{paczynski1967,lipunov1987,hils1990,schneider2001,nelemans2001,ruiter2010,belczynski2010,nissanke2012,korol2017,lamberts2018,lamberts2019,breivik2020,eccentricity_lau}; an overview of these is given in Sect. 1 of \citet{lisa_astrophysics}. The most numerous of these are expected to be WD–WD binaries in the Galactic disk.

We have previously investigated the LISA binary sources in a generic, averaged stellar population in \citet{wouter_gw_spectral}. In this paper, we focus on two different kinds of real, localised stellar populations that exist close to the Galaxy: globular clusters (GCs) and the Magellanic Clouds. The reason for this is that, if they contain LISA-detectable binaries, the detector could potentially resolve these localised populations on the sky against the background of the Galactic disk. LISA could then give insights on the binary population that would not be possible with EM observations alone, such as resolving binaries too faint to be detected in EM. We note that both GCs \citep[e.g.][]{lisa_gc1,ivanova2006,kremer2018_gc_lisa} and the Magellanic Clouds \citep{roebber2020,korol2020,lmclisa} have previously been considered as potential hosts of GW sources for LISA.

\subsection{Globular clusters and Magellanic Clouds}

GCs and the Magellanic Clouds differ in a number of ways, one of which is size: the Magellanic Clouds, being dwarf galaxies, are several orders of magnitude more massive than even the largest GCs. A more subtle distinction is found in the composition and history of the stellar populations they contain: GCs are typically considered to be the result of a single cloud of gas collapsing, resulting in a burst of star formation that creates a population of stars that all have the same age and metallicity. However, this traditional view has been challenged by observations which have shown that some GCs contain multiple subpopulations with distinct chemical compositions; this phenomenon has been theorised to be the result of either some stars in the GCs being polluted by the material ejected from others, or this ejected material coalescing to form a new generation of stars \citep[for an overview, see][and references therein]{bastian2018,milone2022}.

By contrast, the Magellanic Clouds contain stars of many different ages and compositions, produced over an extended period of star formation. We note that despite the existence of multiple populations, GCs are still far more homogeneous than the Magellanic Clouds. For this study, we treat GCs as consisting of a single population, but we take into account the star formation history (SFH) of the Magellanic Clouds using the models of \citet{hz_sfh_smc,hz_sfh_lmc}; this is described in more detail in Sect. \ref{method_chapter}.

Another relevant distinction between these two types of populations is that the Magellanic Clouds have a relatively low stellar density, similar to the Galactic disk, so we can treat their binaries as evolving in isolation, i.e. without external effects upon the evolution. However, GCs have a higher stellar density and correspondingly lower average distance between stars. As a consequence, the chance that a stellar system will have a gravitational encounter with another system, wherein the orbits of the systems are affected or stars are even exchanged, is not negligible. These effects are known collectively as dynamical interactions \citep[see e.g.][]{heggie1975,heggie2003,benacquista2013,kremer2020}; we discuss the effects that dynamical interactions can have on the population of GW sources in a GC in detail in Sect. \ref{globular_discussion}.

We do not simulate dynamical interactions in this paper, but as it is overall not clear from earlier results how strongly dynamical interactions will increase or decrease the number of LISA-detectable binaries in a GC, particularly for WD–WDs (the most numerous GW sources in the LISA band), we consider that studying the GW sources from an isolated binary population still provides a worthwhile comparison. Furthermore, performing these simulations with isolated binaries and then comparing our results to those found using other codes that do include dynamical interactions gives us an indication of the impact dynamical interactions have upon the binary population.

This paper is structured as follows: in Sect. \ref{method_chapter} we present our methods for modelling the GW sources in GCs and the Magellanic Clouds, in Sect. \ref{results_chapter} we present our results, in Sect. \ref{discussion_chapter} we discuss our results and compare these to those of similar studies in the literature and in Sect. \ref{conclusion_chapter} we summarise our conclusions.

\section{Method} \label{method_chapter}
\subsection{Modelling stellar populations with BPASS}

To simulate the stellar populations that we drew GW sources from, we used the Binary Population and Spectral Synthesis (\textsc{bpass}) code suite, which simulates the evolution of a population of binary and single-star systems from a wide range of initial conditions \citep{bpass1,bpass2}. The stellar models that we used are the same as those described in Sect. 2.3 of \citet{wouter_gw_spectral}. To summarise, we use \textsc{bpass} version 2.2.1, with an initial mass function (IMF) based upon \citet{kroupa1993} and initial binary parameters taken from Table 13 of \citet{moe2017}. The \textsc{bpass} population of GW sources is output by the module \textsc{tui} \citep[first discussed in][]{bpassmassdist}.

We use the results output by \textsc{tui} to calculate a population of sources for the appropriate age, metallicity and initial mass. To calculate the expected number of binaries emitting GWs in a given frequency range, we select the binaries in that range from the \textsc{tui} results. Note that this expected number is weighted by the IMF and the initial binary parameters and is thus not an integer.

The frequency range in which we generated binaries for LISA is from $10^{-3.6}$ to $10^0$ Hz, based on our evaluation of the detectability of GW spectra in \citet{wouter_gw_spectral}.

For each iteration of randomly sampling a population of binaries, we used this expected number as the mean of a Poisson distribution from which we obtained an integer number of binaries to generate that would comprise the GW population for that iteration. Subsequently, we selected that number of binaries from the \textsc{bpass} population of the appropriate age and metallicity, with the probability of a given model being selected proportional to its IMF weight; any individual model could be selected more than once. This gave us, for each iteration of the cluster, a set of binary models that would comprise the population of GW sources for that iteration.

\subsection{Stellar population mass calculations} \label{initial_mass_calc}

One important detail of the method we use to sample the populations is that the mass used to calculate the IMF weights needs to be the initial mass of the stellar population; i.e. the mass when the stars were formed. However, what EM observations of GCs give us is the mass of the cluster at the current time, which would be reduced from the initial mass due to material being ejected from the cluster. We accounted for this by computing the initial mass of each GC from its current mass using \textsc{bpass} data files that detail the fraction of surviving stellar mass over time in a stellar population subject to mass loss due to stellar winds \citep{bpass1,bpass2}.

GCs, aside from the aforementioned issue of multiple populations, can generally be treated as having formed in a single starburst and thus being composed of stars that all have roughly the same age and metallicity. However, the Magellanic Clouds underwent star formation over a longer period of time, therefore requiring multiple ages and metallicities to be simulated as well as a different method for calculating the mass. We used the SFH data calculated by \citet{hz_sfh_smc} for the SMC and \citet{hz_sfh_lmc} for the LMC. This data consists of the star formation rates in the Magellanic Clouds binned across different ages (18 bins for the SMC, 16 for the LMC) and metallicities (3 bins for the SMC, 4 for the LMC).

Note that for the LMC there is some ambiguity in the data given by \citet{hz_sfh_lmc} regarding the boundaries of the oldest age bin; their formulation seems to be including star formation prior to the birth of the Universe, but we have excluded that from our bins, so we may be slightly underestimating the star formation compared to \citet{hz_sfh_lmc}. This does not apply for the SMC data given by \citet{hz_sfh_smc}.

We calculated the total stellar mass formed in each age and metallicity bin in the SFH data, giving us the masses of a set of subpopulations, each with a single age and metallicity. We sampled the binaries of these subpopulations as if they were individual populations, and finally combined the generated samples for each subpopulation into a single set of binaries for the entire SMC and LMC each.

Our choice to employ this method of stochastically sampling the \textsc{bpass} models was motivated by earlier research using \textsc{bpass} which showed that stochastically sampling binary parameters, as opposed to simply averaging over all the models in the data set, had an impact upon predictions of various EM-observable properties of stellar populations \citep{bpassstochastic1,bpassstochastic2}.

\subsection{Evaluating GW detectability with LEGWORK}

To evaluate the GW detectability of the binaries in our sampled populations, we used \textsc{legwork}, a software package for making predictions about stellar-origin GW sources and their detectability by space-based GW detectors \citep{legworkjoss,legworksci}.

Using \textsc{legwork}, for each binary in our samples we calculated the signal-to-noise ratio (S/N) relative to the LISA sensitivity curve as given by \citet{robson2019}. We used the fiducial \textsc{legwork} parameters for LISA, aside from the observation time, which we vary between 4 and 10 yr as indicated in the tables. We used a S/N threshold value of 7 to consider a binary to be detectable.

Note that in the \textsc{bpass} output the chirp masses and frequencies of the binaries are binned in logarithmic bins 0.1 dex wide; for the plots shown in the next section we have added random values of up to ±0.05 dex to each of these quantities to avoid multiple systems coinciding visually. We did not do this for the numerical calculations in Sect. \ref{results_chapter} as the effect upon the averaged numbers of detectable binaries is negligible.

\section{Results} \label{results_chapter}

\subsection{GW sources in the Magellanic Clouds} \label{magellanic_results}

\begin{figure*}
    \centering
    \includegraphics[width=1.5\columnwidth]{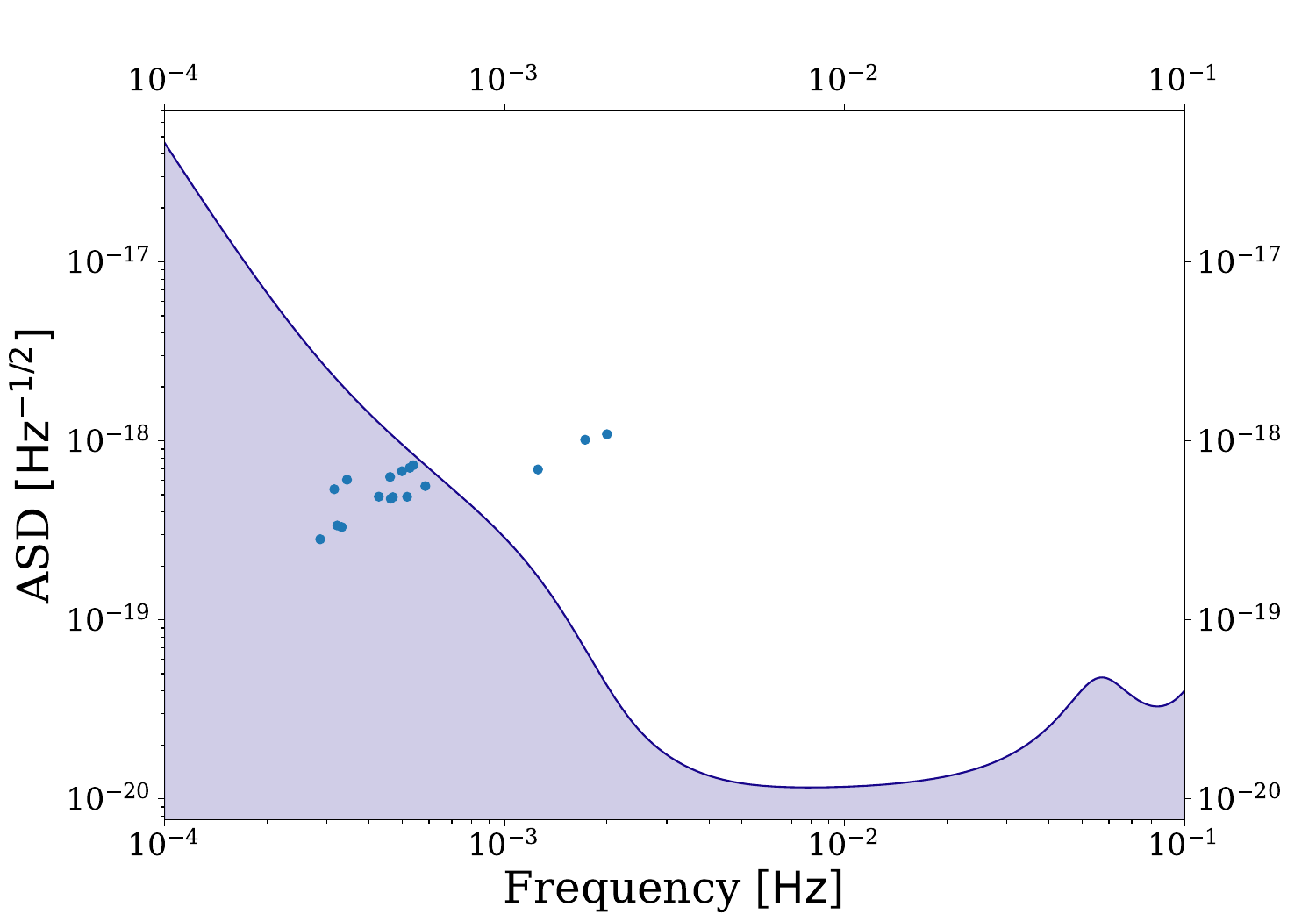}
    \includegraphics[width=1.5\columnwidth]{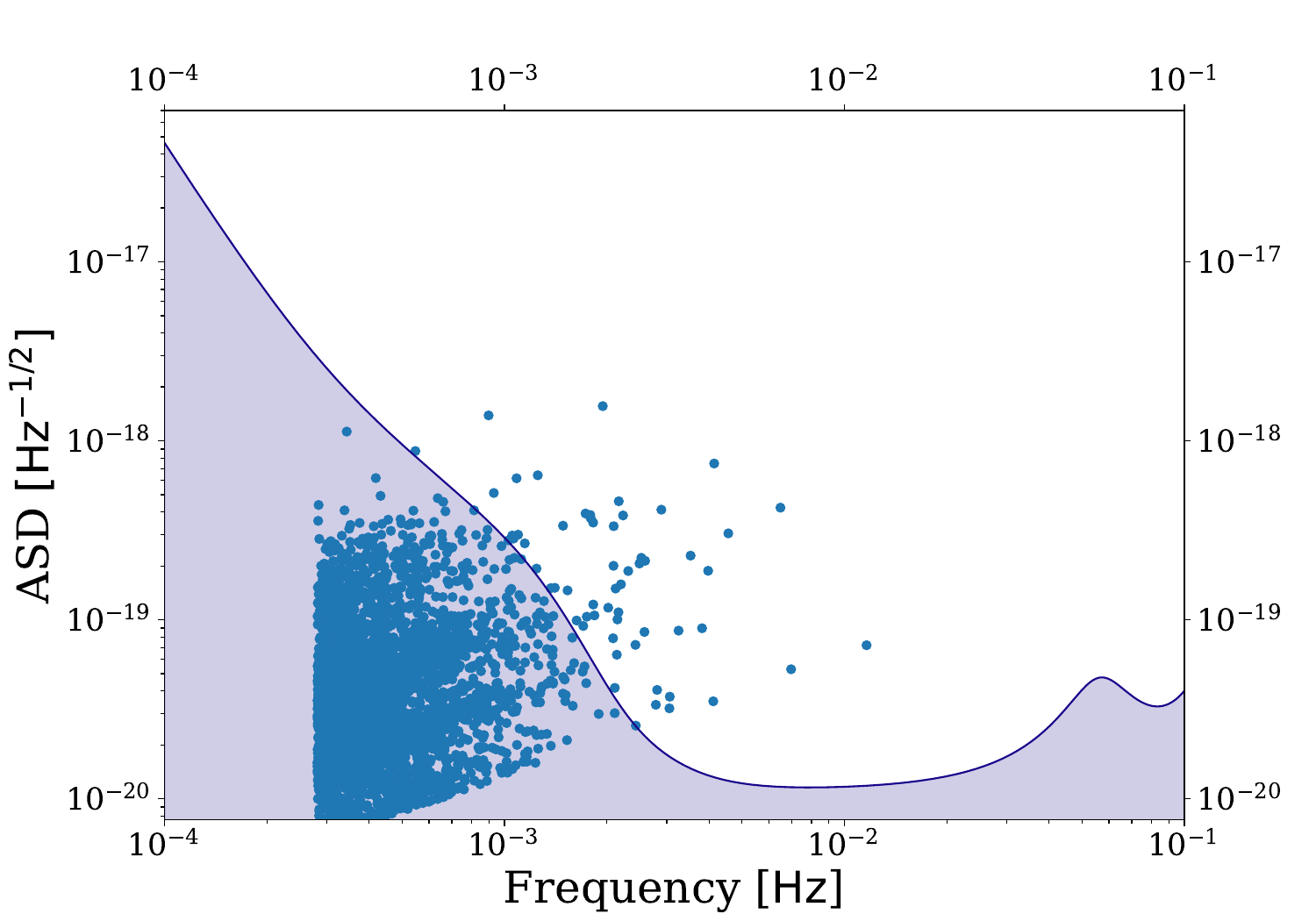}
    \caption{Plots generated using \textsc{legwork} comparing the strain of binaries in two example populations to the LISA sensitivity curve. The first panel is for a population based on Omega Centauri and the second is based on the LMC. Binaries with frequencies below $10^{-3.6}$ Hz are not shown.}
    \label{example_legwork}
\end{figure*}

\begin{table*}
    \centering
    \begin{tabular}{c c | c c c c c c | c}
    Object & $T_{\rm obs}$ (yr) & BH–BH & BH–NS & BH–WD & NS–NS & NS–WD & WD–WD & Total \\
    \hline
    SMC & 4 & 0.018 & 0.135 & 0.571 & 0.167 & 1.623 & 1.069 & 3.583 \\
    SMC & 10 & 0.019 & 0.189 & 0.834 & 0.199 & 2.182 & 2.252 & 5.675 \\
    LMC & 4 & 0.026 & 0.358 & 1.570 & 0.395 & 3.965 & 4.878 & 11.192 \\
    LMC & 10 & 0.045 & 0.528 & 2.014 & 0.591 & 5.190 & 9.384 & 17.752 \\

    \end{tabular}
    \caption{Average numbers of detectable binaries for 4 and 10 years of LISA observations from 1000 simulations of the SMC and LMC. The standard deviation from Poisson uncertainty is 0.032 for all values. Binaries are considered detectable if their S/N > 7.}
    \label{magellanic_table}
\end{table*}

Fig. \ref{example_legwork} shows two plots of example populations generated using \textsc{legwork}, to illustrate both the types of populations we work with in this paper and the outputs of \textsc{legwork}. We plot the strain (in terms of amplitude spectral density) and frequency of each binary in the population and compare these to the LISA sensitivity curve as given by \citet{robson2019}. The first plot shows a population with parameters based on the GC Omega Centauri, using data we discuss in Sect. \ref{globular_results}. The second plot shows a simulation of the LMC using the SFH data of \citet{hz_sfh_lmc}.

In the Omega Centauri example population shown in Fig. \ref{example_legwork}, there are 17 sources in the LISA frequency range; of these three lie above the sensitivity curve, with two having a S/N > 7. In the LMC example population, there are over 5000 binaries. 
The vast majority of the binaries in this example lie below the sensitivity curve even with the frequency cutoff we have applied.

Moving on to the first part of our main results, Table \ref{magellanic_table} shows the results of 1000 random samplings of the SMC and LMC using the SFH data of \citet{hz_sfh_smc} and \citet{hz_sfh_lmc}, respectively; each of these samples are individually similar to that in the bottom panel of Fig. \ref{example_legwork}. The table shows the average numbers of detectable binaries of each compact binary type in the Magellanic Clouds. We show the amount of detectable binaries after both 4 and 10 years of observation.

After 10 years of observation, we expect around 6 detectable binaries for the SMC and 18 for the LMC. For both 4 and 10 years of observation, we find around three times as many detectable sources for the LMC as for the SMC. For both the SMC and LMC, the number of detectable binaries increases by around 60\% going from 4 to 10 years of observation.

For the SMC, the most common types of detectable binaries are NS–WDs followed by WD–WDs after 4 years of observation, and after 10 years of observation WD–WDs become the most common. For the LMC the most common types are WD–WDs followed by NS–WDs for both 4 and 10 years of observation. For both Magellanic Clouds, the third most common type is BH–WDs, which have an expected number greater than one for the LMC but not for the SMC. NS–NSs and BH–NSs are less common, but still have expected numbers greater than 0.5 each for the LMC after 10 years of observation. BH–BHs are the least likely type of compact binaries to be detected. This hierarchy generally matches with what we found for the numbers of binaries in the LISA frequency range in Table 1 of \citet{wouter_gw_spectral}, though the relative differences between the most common and least common types are smaller here, likely because the less commonly occurring types of binaries are precisely those which are more massive and thus individually easier to detect in GWs if they are in fact present.

\subsection{GW sources in globular clusters} \label{globular_results}

\begin{table*}
    \centering
    \begin{tabular}{c c c | c c}
    Object & Distance (kpc) & M$_{\rm current}$ (M$_{\odot}$) & No. binaries, 4 yr & No. binaries, 10 yr \\
    \hline
    $\omega$ Cen & 5.43 & $3.64 \times 10^6$ & 0.2359 & 0.4316 \\
    Terzan 5 & 6.62 & $9.35 \times 10^5$ & 0.0594 & 0.0764 \\
    Liller 1 & 8.06 & $9.15 \times 10^5$ & 0.0572 & 0.0708 \\
    47 Tuc & 4.52 & $8.95 \times 10^5$ & 0.0720 & 0.1071 \\
    
    \end{tabular}
    \caption{Average numbers of detectable binaries for 4 and 10 years of LISA observations from 100,000 simulations of several globular clusters. The standard deviation from Poisson uncertainty is 0.0032 for all values. The physical parameters were taken from \citet{catalogue_baumgardt1} and \citet{catalogue_baumgardt2} and compiled in Table C2 of \citet{wouter_gw_spectral}.}
    \label{gc_table}
\end{table*}

Table \ref{gc_table} shows the results of 100,000 iterations of four different GCs; each of these samples are individually similar to that in the upper panel of Fig. \ref{example_legwork}. The physical parameters of these were taken from \citet{catalogue_baumgardt1} and \citet{catalogue_baumgardt2} and compiled in Table C2 of \citet{wouter_gw_spectral}, with the metallicity taken to be Z = 0.001. These GCs were chosen as the four most massive GCs in the catalogue located within 10 kpc of Earth.

We expect that each of these GCs will have at least several binaries emitting GWs in the LISA frequency band based on the findings in \citet{wouter_gw_spectral}, but Table \ref{gc_table} shows that the amount of binaries that would be detectable by LISA is lower. Specifically, comparing the results in Table \ref{gc_table} to the total numbers of binaries in Table C2 of \citet{wouter_gw_spectral} suggests that only two to three percent of the binaries emitting in the LISA frequency band in these GCs would actually be individually detectable after 10 years of LISA observation.

For Omega Centauri, the most massive GC orbiting the MW, the expected number of detectable binaries is smaller than one even after 10 years of observation, and for the other GCs – which are still amongst the most massive in the MW – the probability of a detectable binary being present is ten percent or less.

The four GCs we investigate are all between $10^{10}$ and $10^{10.1}$ years old, so we can compare these to the \textsc{bpass} data on different types of compact remnants in Table 1 of \citet{wouter_gw_spectral}, specifically to the row of the table for a population $10^{10}$ years old with metallicity Z = 0.001. Based on that data, we can surmise that any binaries that do appear in our simulations of these GCs have a 93\% chance of being a WD–WD, with the next most likely types being BH–WDs followed by NS–WDs. Irrespective of the metallicity, WD–WDs are by far the most common type of compact binaries in \textsc{bpass} stellar populations of this age.

\section{Discussion} \label{discussion_chapter}

\subsection{GW sources in the Magellanic Clouds} \label{magellanic_discussion}

\subsubsection{Comparison to previous research}

The LMC results in Sect. \ref{magellanic_results} can be compared to those of \citet{lmclisa}, who also evaluate the number of GW sources that would be detectable by LISA in the LMC using the SFH of \citet{hz_sfh_lmc}, but instead of \textsc{bpass} use \textsc{SeBa} stellar evolution/population synthesis models \citep{portegies1996,nelemans_seba,seba_toonen}.

\citet{lmclisa} predict around 100 to several hundred WD–WD binaries that would be detectable by LISA after 4 yr of observations, depending on assumptions on the stellar distribution. This quantity is at least a factor of 20 higher than our corresponding value in Table \ref{magellanic_table}. \citet{lmclisa} also predict a total of 1–2 $\times 10^6$ WD–WD binaries emitting with a frequency of at least $10^{-4}$ Hz in the LMC (regardless of detectability), where for the same frequency range (which is somewhat wider than our default values for the LISA frequency range) we find around 3–4 $\times 10^4$ binaries of any type.

Notably, \citet{bpassmilkyway}, using \textsc{bpass} to simulate the WD–WD population of the Milky Way, found a similar discrepancy in the amounts of LISA-detectable WD–WD binaries to what we find here; they found approx. a factor of 20 fewer of these than a comparable study done by \citet{lamberts2018,lamberts2019} using the stellar evolution/population synthesis code \textsc{bse} \citep{hurley2002}.

Another difference between our work and \citet{lmclisa} is that they only looked at WD–WD binaries in their study of LISA GW sources in the LMC, while our data in Table \ref{magellanic_table} predicts at least one detectable NS–WD and BH–WD to be present in the LMC, and non-negligible chances of detectable NS–NS and BH–NS also. This indicates that it is also relevant to study non-WD–WD binaries as potential LISA sources in the LMC.

\subsubsection{Mass transfer and CEE in stellar evolution codes}

As the discrepancy in the amounts of LISA-detectable WD–WD binaries appears in two separate studies, we believe it is caused by an intrinsic difference in the modelling of stellar evolution between \textsc{bpass} on the one hand and \textsc{SeBa} and \textsc{bse} on the other. A distinction can be drawn between these two codes in that \textsc{bpass} models stellar structure in a detailed way throughout the evolution while \textsc{SeBa} and \textsc{bse} use pre-computed stellar tracks, which do not achieve the same amount of detail in stellar structure but are computationally faster, thus allowing a greater number of different initial parameters to be investigated. However, we note that all stellar evolution models, including detailed ones like \textsc{bpass}, depend on many assumptions and simplifications about stellar evolution and binary interactions, and in this case it is likely that these assumptions are the most important factor leading to the differing results.

More specifically, we think that the difference in the numbers of detectable WD–WD binaries relates to the treatment of common-envelope events (CEEs) and the stability of mass transfer in the different codes. There have been several recent studies focusing on the implementation of mass transfer in \textsc{bpass} \citep{bpassmasstransfer,bpasssuperedd}, which have found that in \textsc{bpass} mass transfer is more likely to be stable than in other stellar evolution codes used for population synthesis, and correspondingly fewer common-envelope events occur. This relates to the different criteria used in stellar evolution codes to determine mass transfer stability: in \textsc{bpass} a binary is considered to have entered CEE when the radius of the donor (the star that is transferring material to its companion) exceeds the separation between the cores of the two stars \citep{bpassgw170817}, while rapid codes such as \textsc{bse} use analytic calculations based upon the evolutionary state of the donor and the binary's mass ratio to determine stability \citep{hurley2002}. More detail on how \textsc{bpass} determines mass transfer stability compared to \textsc{bse} can be found in Sect. 2.3.2 of \citet{bpassmilkyway} and Sect. SI.2.4 of \citet{bpassgw170817}.

Though \citet{bpassmasstransfer} and \citet{bpasssuperedd} have focused on binaries that produce BHs and NSs rather than the lighter stars that would produce WDs, it has also been found that in \textsc{bpass} no CEEs occur at all for binaries with masses below 2 M$_{\odot}$ \citep{bpassmilkyway}. The prescription used in \textsc{bpass} for CEEs themselves is also quantitatively different from those used in other codes, as \textsc{bpass} performs step-by-step calculations of the stellar structure throughout the CEE, and on average tends to lead to less orbital shrinkage than in other codes \citep{bpassgw170817}, which also affects the resulting period distribution of compact binaries.

The two differences that we have described between \textsc{bpass} and other codes – mass transfer being more likely to remain stable rather than forming a CEE, and less angular momentum being lost in CEEs – could in different circumstances either decrease or increase the number of LISA-detectable binaries. Which of these cases applies depends on how efficient the CEE in the simulation is to begin with. In either case, increasing the efficiency of CEE reduces the amount of angular momentum loss and increases the average separation of the binary after the CEE. Relatively low efficiency means that for many binaries there is so much angular momentum loss that the binary merges before both stars have become compact objects. In this regime, increasing the CEE efficiency increases the amount of LISA-detectable binaries by reducing the amount of early mergers, as found by e.g. \citet{thiele2023}. However, if the CEE efficiency is much higher, then many binaries will remain so wide after the CEE that their GW frequency is too low for LISA to detect, and increasing the CEE efficiency in this regime decreases the amount of LISA-detectable binaries. As we detail in the next subsection, the CEE of \textsc{bpass} is much more efficient than that of most other population synthesis codes, and so \textsc{bpass} operates in the second of these regimes while other simulations like those of \citet{thiele2023} are in the first.

\subsubsection{Details on mass transfer and CEE in BPASS}

\begin{figure*}
    \centering
    \includegraphics[width=1.5\columnwidth]{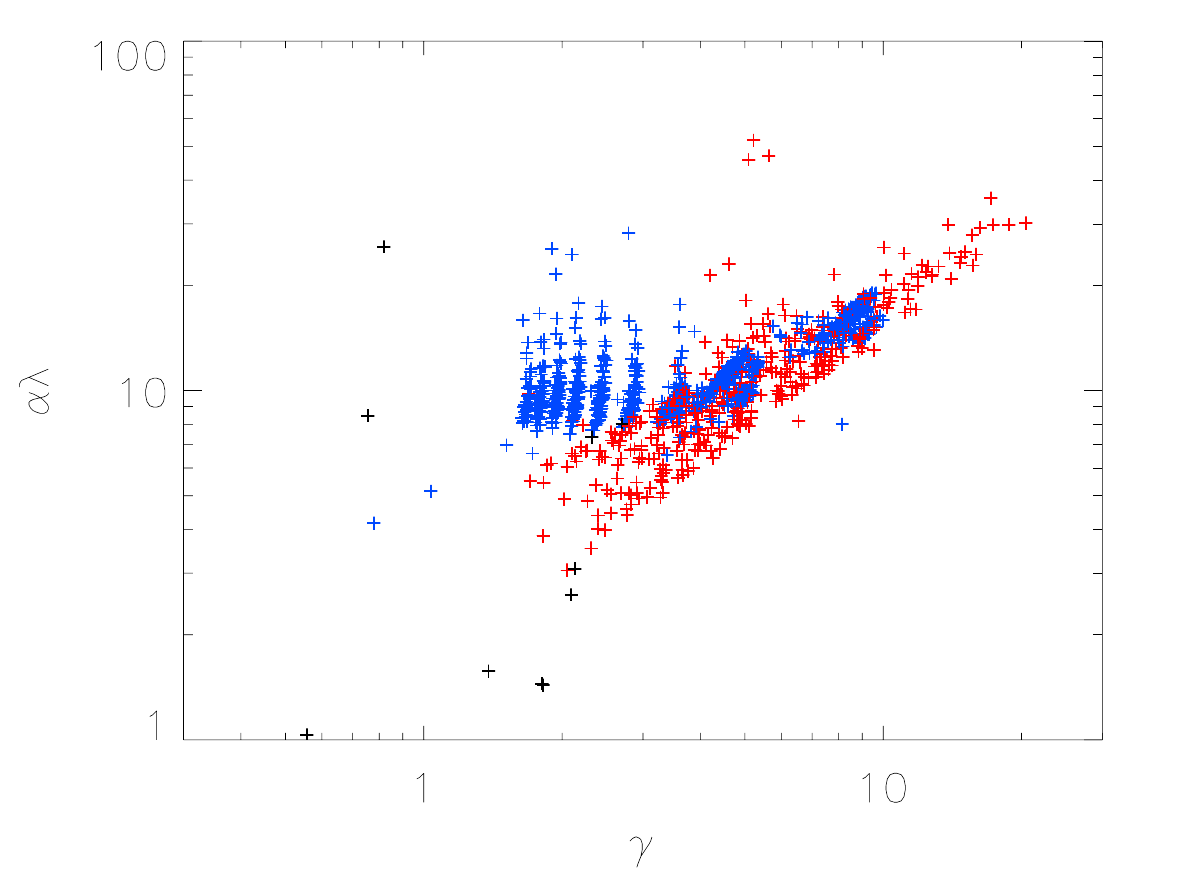}
    \caption{The effective $\alpha \lambda$ and $\gamma$ CEE prescription parameters for \textsc{bpass} models that experience CEE. The black points are systems that merge in the first binary interaction, the blue points are CEEs from the first binary interaction of a system and the red points are CEEs from the second binary interaction where one of the stars is a WD.}
    \label{bpass_cee_plot}
\end{figure*}

To fully understand why \textsc{bpass} predicts fewer LISA-detectable WD–WDs would require a model-by-model comparison between the \textsc{bpass} stellar evolution models and those from a rapid population synthesis code. Such a detailed comparison is beyond the scope of this paper; however, we are able to gain some insights by looking at some of the details of the \textsc{bpass} models.

At solar metallicity, for a simple stellar population with a total mass of $10^6$~M$_{\odot}$, for the first binary interaction with the primary star as the donor, \textsc{bpass} predicts there would be 20,907 systems experiencing stable mass transfer, 6783 experiencing CEE and 223 merging. Then, for the second binary interaction when the donor is the original secondary star, 2361 systems experience stable mass transfer but 7349 experience CEE.

From these numbers it is clear that in \textsc{bpass} the first binary interaction is stable \citep[this can also be seen in Fig. 1 of][]{bpass1}. In the \textsc{bpass} stellar evolution models, mass transfer stability is determined by how the stellar model response to Roche lobe overflow (RLOF). In stable mass transfer the mass loss in RLOF prevents the growth of the stellar radius, which prevents a CEE. However, CEE can occur through the Darwin instability or if the star continues to grow despite the onset of RLOF. CEE is assumed to occur when the radius of the stellar model is greater than the binary separation.

In addition to the stability of mass transfer, the second factor affecting our WD–WD period distribution is that the \textsc{bpass} CEE model is efficient, i.e. it leads to very weak shrinking in the binary orbit during CEE. To demonstrate this, we include estimated effective values for our CEE models of the $\alpha \lambda$ and $\gamma$ values in Fig. \ref{bpass_cee_plot}. We see that the effective $\alpha \lambda$ values are between 8 to 20 for CEE in the first binary interaction. Lower values are possible for the second binary interaction, with values from 4 to 30. These values are much higher than those typically used in other stellar evolution codes; for example, the \textsc{SeBa} version of \citet{seba_toonen} assumes $\alpha \lambda = 2$ and $\gamma = 1.75$, based upon calculations by \citet{nelemans_gamma1} and \citet{nelemans_seba}, respectively. We suspect this is the primary reason why our WD–WD binary population has a dearth of low-period binaries compared to other predictions.

The final two relevant features of the \textsc{bpass} CEE are that, firstly, the stellar models are computed in a detailed code. With this code it is difficult to just remove the entire envelope in a single timestep as is normally assumed in rapid population synthesis models. Instead, we remove the mass as at high a rate as possible until the envelope collapses and the star is once more smaller than its Roche lobe. The fact this takes time leads to a small spread of possible CEE parameter values, because they are dependent on the stellar and binary parameters. The second feature is that the prescription is closer to the $\gamma$-prescription \citep{nelemans_gamma1,nelemans_gamma2} in which angular momentum is conserved, rather than the $\alpha$-prescription \citep{webbink1984} which conserves energy; this has also been discussed in \citet{bpassgw170817} for its implications on the merger rate of NS binaries.

In summary, the treatment of binary interactions in \textsc{bpass} is inherently different to most other stellar evolution models. The greater stability of the first binary interaction and the more efficient CEE prescription together lead to fewer short-period WD–WD systems for LISA to detect.

\subsubsection{Other studies and additional discussion on mass transfer assumptions}

\citet{temmink2023} investigated mass transfer stability using the detailed stellar evolution code \textsc{mesa} \citep{mesa_2011,mesa_2019}, and similarly to the \textsc{bpass} research found that mass transfer in their simulation tended to be more stable than under the assumptions used in other codes such as \textsc{bse}, but that still a significant fraction of binaries become unstable and likely go through a CEE.

Additionally, there has also been recent research using \textsc{SeBa} on the effect of changing assumptions about mass transfer upon the resultant population of GW sources \citep{seba_masstransfer}, though like \citet{bpassmasstransfer} this research only focused on BH binaries.

Similarly, \citet{eccentricity_wagg} performed a metastudy (in their Fig. 12) comparing LISA GW detection rate predictions between different stellar evolution codes, showing that these varied by more than an order of magnitude, which is comparable to the differences between codes we have found. Separately, \citet{eccentricity_wagg} also created their own models which used \textsc{legwork} and the rapid population synthesis code \textsc{compas} \citep{compas_1,compas_2,compas_3}, and when they varied physics assumptions in these they saw changes of similar magnitude to those we found, though they also only focused on binaries with BHs and NSs.

Some more information on the differences in detectable WD–WD binaries can be found in Fig. 7 of \citet{bpassmilkyway}, which compares the WD–WD distribution in chirp mass and period between the \textsc{bpass} and \textsc{bse} galaxy models. The \textsc{bse} distribution shows a trail of binaries with low chirp masses (below 0.3 M$_{\odot}$) that extends to high frequencies (above $10^{-4}$ Hz) which is not present in the \textsc{bpass} distribution. The diagonal shape of this trail, rather than horizontal, suggests that these high-frequency binaries are not simply WD–WDs formed at lower frequencies evolving solely through GW emission, but rather that those binaries must have formed at those higher frequencies, which suggests they must have undergone some form of CEE. A similar diagonal trail of low-chirp-mass, high-frequency WD–WDs can also be seen in the \textsc{SeBa} results in Fig. 2 of \citet{nelemans_seba}.

We note that there are several tens of short-period WD–WDs known from EM observations which would have GW frequencies between $10^{-4}$ and $10^{-3}$ Hz; see e.g. Fig. 1 of \citet{seba_toonen} or the more recent observations catalogued by \citet{ztf_wdwd_cat}. In general, earlier research with \textsc{SeBa} has shown that its results for WD–WD binaries agree quite well with EM observations in terms of the mass and period distribution of these systems, though these EM observations are subject to observational biases \citep{nelemans_seba,seba_toonen}.

Aside from differences in CEEs and mass transfer, \textsc{bpass} also does not include magnetic wind braking \citep{magnetic_wind_braking}, an effect which can cause binaries to lose angular momentum. By contrast, \textsc{SeBa} \citep{portegies1996} and \textsc{bse} \citep{hurley2002} do include this effect in their models. However, we do not expect this effect to be as important as the common-envelope prescription for the WD–WD binaries observable by LISA.

\subsubsection{Effects of a stronger common-envelope prescription}

\begin{table*}
    \centering
    \begin{tabular}{c c | c c c c c c | c}
    Model & $T_{\rm obs}$ (yr) & BH–BH & BH–NS & BH–WD & NS–NS & NS–WD & WD–WD & Total \\
    \hline
    Fiducial & 4 & 0.026 & 0.358 & 1.570 & 0.395 & 3.965 & 4.878 & 11.192 \\
    Fiducial & 10 & 0.045 & 0.528 & 2.014 & 0.591 & 5.190 & 9.384 & 17.752 \\
    \hline
    A (10\%, 10$\times$) & 4 & 0.225 & 9.169 & 58.193 & 11.109 & 144.78 & 269.84 & 493.31 \\
    A (10\%, 10$\times$) & 10 & 0.244 & 10.901 & 75.287 & 15.558 & 183.09 & 526.68 & 811.77 \\
    \hline
    B (20\%, 10$\times$) & 4 & 0.483 & 17.815 & 114.88 & 21.604 & 286.76 & 535.98 & 977.52 \\
    B (20\%, 10$\times$) & 10 & 0.503 & 20.973 & 148.80 & 30.448 & 361.86 & 1043.9 & 1606.5 \\
    \hline
    C (100\%, 3$\times$) & 4 & 0.476 & 5.477 & 28.868 & 5.967 & 84.176 & 108.08 & 233.04 \\
    C (100\%, 3$\times$) & 10 & 0.653 & 7.655 & 37.270 & 7.189 & 98.546 & 170.92 & 322.23 \\

    \end{tabular}
    \caption{Average numbers of detectable binaries for 4 and 10 years of LISA observations from 1000 simulations of the LMC, including our fiducial model and several simple models of a population with more orbital shrinkage from mass transfer and CEE. The standard deviation from Poisson uncertainty is 0.032 for all values. Binaries are considered detectable if their S/N > 7.}
    \label{cee_table}
\end{table*}

To investigate the effects that differences in CEE prescriptions could have on the amounts of binaries detectable by LISA, we make several kinds of simple modifications to our \textsc{bpass} populations to simulate a stronger CEE prescription (i.e. one that produces more orbital shrinkage) than that of \textsc{bpass}. Table \ref{cee_table} shows the results of 1000 random samplings of the LMC, including both the fiducial model (identical to that shown in Table \ref{magellanic_table}) and three adjusted models.

In each of the three adjusted models, the orbital frequency of some of the binaries in the population is increased, simulating a situation in which a CEE led to stronger orbital shrinkage for those binaries. In Model A, we randomly select 10\% of the binaries in the \textsc{bpass} output population and increase their frequency by a factor of 10. In Model B, we do the same except we randomly select 20\% of the binaries instead. In Model C, we increase the frequency of all of the binaries by a factor of 3.

In Table \ref{cee_table}, we can see that in the alternate models, even those where we only adjust 10\% of the binaries in the population, the number of binaries detectable by LISA is dramatically increased. For Model A, there are around 45 times as many detectable binaries in total compared to the fiducial model, and for Models B and C the factors of increase are around 90 and 20, respectively. The factors are similar after both 4 and 10 years of observation.

For Model C (with a threefold frequency increase) the factor by which the amount of detectable binaries is increased is fairly consistent across each of the binary types, but for Models A and B (with a tenfold frequency increase) the lighter binary types are increased more than the heavier binary types, which gives an indication of differences in the underlying period distributions of the different binary types, particularly at frequencies around $10^{-4}$ and $10^{-3}$ Hz where binaries may not be detectable in the fiducial model but would be with a tenfold frequency increase. BH–BHs are increased by a factor of around 10 in Model A and 20 in Model B, while WD–WDs are increased by a factor of around 50 in Model A and 110 in Model B.

The key result of this test is that amounts of detectable WD–WD binaries predicted by the alternate models, particularly Models A and B, fall within the range of the \textsc{SeBa} results of \citet{lmclisa}, who predicted from one hundred to several hundred detectable WD–WDs (depending on assumptions in their models) after four years of LISA observation. This gives us further confidence that the main factor leading to the difference between our fiducial results and those of \citet{lmclisa} is the treatment of mass transfer and CEE in \textsc{bpass} vs. \textsc{SeBa}, as with these simple approximations of stronger CEE we can make the \textsc{bpass} results match those from \textsc{SeBa}.

However, we note that these implementations of stronger CEE are highly simplified and there are a number of caveats with these. For example, for Models A and B we select the binaries that we apply the frequency increase to randomly from across all binary types, while in reality the effects of CEE would depend on the properties of the binary and be different for different binary types. Additionally, the three alternate models do not take into account that the increase in frequency from stronger CEE would cause the binaries to merge more quickly through emission of GWs, which would affect the age distribution of these binaries.

\subsubsection{Effects of confusion noise models}

\begin{table*}
    \centering
    \begin{tabular}{c c | c c c c c c | c}
    Object & $T_{\rm obs}$ (yr) & BH–BH & BH–NS & BH–WD & NS–NS & NS–WD & WD–WD & Total \\
    \hline
    SMC & 4 & 0.021 & 0.231 & 0.857 & 0.262 & 2.063 & 1.384 & 4.818 \\
    SMC & 10 & 0.030 & 0.419 & 1.311 & 0.329 & 3.100 & 2.252 & 7.441 \\
    LMC & 4 & 0.051 & 0.646 & 2.352 & 0.703 & 4.821 & 4.878 & 13.451 \\
    LMC & 10 & 0.066 & 1.146 & 3.385 & 1.043 & 8.601 & 10.967 & 25.208 \\

    \end{tabular}
    \caption{Average numbers of detectable binaries for 4 and 10 years of LISA observations from 1000 simulations of the SMC and LMC, as in Table \ref{magellanic_table}, except that the confusion noise used in the S/N calculations is set to zero.}
    \label{magellanic_confusion_table}
\end{table*}

In our detectability calculations, we use the LISA sensitivity curve constructed by \citet{robson2019}. This curve gives the strain sensitivity of LISA at different frequencies, and consists of two components: the instrumental noise (noise originating from physical properties of the detector itself) and foreground confusion noise, which consists of the overlapping signals of many Galactic binaries (mostly WD–WDs) which are individually too weak to resolve. This noise, therefore, depends on the population of binary sources present in the Galaxy.

The sensitivity curve of \citet{robson2019} includes a formulation of the confusion noise from \citet{cornish2017}, who calculated the confusion noise using a galaxy model constructed using \textsc{SeBa}. As established in the preceding sections of this paper, stellar populations modelled using \textsc{bpass} yield fewer binaries in the LISA frequency range compared to \textsc{SeBa}. Therefore, it is not entirely accurate for us to be calculating the detectability of \textsc{bpass} stellar populations against a sensitivity curve which was calculated using a \textsc{SeBa}. This is because we would expect the confusion noise to be smaller in magnitude if it was calculated using a \textsc{bpass} galaxy model like that of \citet{bpassmilkyway}, as the binaries contributing to the unresolved Galactic foreground would be fewer in number.

With less confusion noise, the number of binaries that would be detectable by LISA would increase. Constructing a new LISA sensitivity curve using \textsc{bpass} is beyond the scope of this paper, but we can calculate the detectability in a maximally optimistic case by simply assuming the confusion noise to be zero. The results of this are shown in Table \ref{magellanic_confusion_table}, where we repeat our calculations of the number of detectable binaries in the LMC and SMC (as in Table \ref{magellanic_table}) but with the confusion noise set to zero in \textsc{legwork}. The values that would be found using \textsc{bpass}-based confusion noise must then be between those using \textsc{SeBa}-based confusion noise in Table \ref{magellanic_table} as a lower bound and those using zero confusion noise in Table \ref{magellanic_confusion_table} as an upper bound.

As expected, the numbers of detectable binaries in Table \ref{magellanic_confusion_table} are higher than those in the fiducial case for both the SMC and the LMC and for both 4 and 10 years of observation. However, the total number of binaries increases by a factor of only approx. 30\% on average, and in none of the cases does the increase exceed 50\%. Looking at WD–WDs alone, the factor of increase is even smaller. Therefore, even in this most optimistic case, altering the confusion noise does not come close to resolving the differences between our results and those of \citet{lmclisa}. Hence, we conclude that recalculating the \citet{robson2019} noise curve to use a \textsc{bpass}-based confusion noise would not affect our previous conclusions.

\subsection{GW sources in globular clusters} \label{globular_discussion}

\subsubsection{Comparison to previous research}

We can compare our GC results in Sect. \ref{globular_results} to the results of \citet{kremer2018_gc_lisa}, who used Cluster Monte Carlo \citep[\textsc{cmc};][]{joshi2000,rodriguez2022} simulations for the gravitational evolution of a GC along with the stellar evolution codes \textsc{bse} and its single-star evolution counterpart \textsc{sse} \citep{hurley2000} to predict that around 14 to 21 binaries across all MW GCs would be detectable by LISA after 4 years, including 4 to 6 WD–WDs. An earlier study by \citet{willems2007}, who used GC simulations by \citet{ivanova2006} using \textsc{cmc} and the stellar evolution code \textsc{startrack} \citep{startrack_1,startrack_2}, found even higher numbers, specifically finding several dozen observable WD–WDs across all MW GCs, even with the constraint of only looking at eccentric binaries.

We have not performed an equivalent computation for all MW GCs, particularly because for less massive GCs the uncertainty from the limited number of iterations would dominate the actual rates. However, based upon our results in Table \ref{gc_table} using some of the most massive GCs, we can say that our total number of expected binaries for all $\sim$150 known MW GCs, particularly for only 4 years of observation, would likely be on the order of one or fewer. Therefore, as with the Magellanic Clouds, we see a lower number of detectable binaries in our simulations compared to previous simulations using different methods.

The size of this discrepancy, a factor of a few tens, is also comparable to that we found for the Magellanic Clouds. This means that, if we account for the differences in stellar evolution between \textsc{bpass} and other stellar evolution codes like \textsc{bse} which we discussed in Sect. \ref{magellanic_discussion}, our numbers would become similar to those of \citet{kremer2018_gc_lisa} and \citet{willems2007}.

\subsubsection{Dynamical interactions}

As for the Magellanic Clouds, the intrinsic differences between \textsc{bpass} and other population synthesis codes in terms of stellar evolution are a probable explanation for our expected numbers of detectable binaries in GCs being lower than those found by other groups. Because the difference between our results and those from other stellar evolution codes is similar in magnitude for both GCs and the Magellanic Clouds, it is possible that the stellar evolution is also the largest contributing factor for GCs. However, for our simulations of GCs there is an additional factor that does not apply for the Magellanic Clouds which could also affect the binary population and the number of detectable GW sources: dynamical interactions.

To be more specific, the \textsc{bpass} models that we use consist of isolated binaries, wherein the stellar evolution is only affected by processes inside the binary, without any effects from external forces. For stellar environments with a relatively low density, such as the MW disk of \citet{bpassmilkyway} or the Magellanic Clouds discussed in Sect. \ref{magellanic_discussion} of this paper, the assumption of isolated binary evolution is reasonable. However, in GCs the stellar density is sufficiently high that the chance of a binary interacting with another stellar system during its evolution is non-negligible.

These dynamical interactions have been studied through simulations of the gravitational evolution of a GC, for which several different methods have been used in the literature. The most accurate method is to fully model the gravitational effect that each star or compact remnant in the cluster has on every other one using an N-body simulation. A number of different codes exist which perform this, including the \textsc{nbody} family of codes, which have a long history of development starting from \textsc{nbody1} \citep{nbody1}, with the most recent version being \textsc{nbody7} \citep{nbody7}.

However, N-body simulations have a significant limitation in the high amount of computational time they require. This prompted the development of another type of GC dynamical simulations, which approximate the N-body simulations numerically with a significantly shorter computational time, particularly the Monte Carlo method which was originally developed by \citet{henon1971a,henon1971b}. There are currently two commonly used computational implementations of this method, which are the aforementioned \textsc{cmc} \citep{joshi2000,rodriguez2022} and \textsc{mocca} \citep{giersz1998,hypki2013}.

\subsubsection{Effects of dynamical interactions on binaries}

Dynamical interactions can have various effects upon the binary population and consequently the GW signal of the GC. It is expected the cumulative effects of many minor interactions will result in a general hardening of the binary population, i.e. an increase in average frequency \citep{heggie1975,ivanova2006,ivanova2008,portegies2010}. Such an effect would tend to shift more binaries into the LISA frequency range, increasing the number of those that would be detectable, and therefore this could be a contributing factor to the differences between our results and those of \citet{kremer2018_gc_lisa}.

However, dynamical interactions could also result in both the formation of new binaries through tidal capture and the unbinding of existing ones. The relative importance of these processes varies depending on the types of binaries considered. Binaries with BHs and NSs that occur in GCs are predicted to mostly be dynamically formed \citep[e.g.][]{kundu2002,pooley2003,sadowski2008,ivanova2008,morscher2015}. This is corroborated when we compare the \textsc{bpass} results to those from \textsc{cmc} simulations: in Table 1 of \citet{kremer2018_gc_lisa} the predictions for detectable BH–BH and NS–NS binaries range from being the same those for WD–WDs to a factor of 10 lower, but in Table 1 of \citet{wouter_gw_spectral} the differences are larger: the NS–NS rates are two to four orders of magnitude lower than those of WD–WDs, and BH–BHs around five orders of magnitude lower than WD–WDs.

The effect of dynamical interactions on WD–WD evolution and rates is less clear from the literature. This is partly because of differing results from different studies and partly because many of the studies on this topic do not directly look at WD–WD binaries, but rather at cataclysmic variables (CVs), particularly those of the AM Canum Venaticorum (AM CVn) type; these are mass-transferring binaries with a WD and a living star, which are easier to detect in EM than WD–WDs. While CVs are not generally thought to evolve directly to GW-emitting WD–WD binaries, they originate from similar types of stellar systems, and therefore the population of EM-observable CVs in a cluster can be indicative of its population of GW-observable WD–WDs.

For example, based upon X-ray observations of CVs in GCs, different analyses by \citet{pooley2006} and \citet{cheng2018_1} reached different conclusions, with the former suggesting that most CVs in GCs were of dynamical origin and the latter the opposite, i.e. that most were of primordial origin; both analyses have potential caveats due to observational limits and biases \citep{heinke2020,belloni2021}. A recent study by \citet{bao2023}, using X-ray observations to study the period distribution of CVs in the GC 47 Tucanae, did find evidence that a significant fraction of these CVs were likely dynamically formed.

In terms of theoretical predictions based upon simulations, \citet{knigge2012} argues based upon a review of a number of different simulations that most CVs in GCs are likely dynamically formed. Similarly, studies using \textsc{cmc} \citep{ivanova2006} and \textsc{mocca} \citep{belloni2016} did find that respectively $60\%$ and $100\%$ of the surviving CV models in their simulations had undergone some sort of dynamical interaction. However, this does not necessarily mean that dynamical interactions increase the number of CVs in a cluster, as \citet{belloni2019} and \citet{kremer2020}, using the \textsc{mocca} and \textsc{cmc} types of Monte Carlo simulations, respectively, found that dynamical destruction of CVs in GCs was stronger than dynamical creation. \citet{kremer2020} also found that the amount of WD–WD binaries in a GC decreases with increasing dynamical interactions. Finally, we note that \citet{shara2006}, using N-body simulations, found 5–6 times more short-period WD–WDs in a GC population compared to the field but 2–3 times \textit{fewer} CVs, which highlights that trends in EM-observable CVs do not necessarily correlate with those in GW-observable WD–WDs.

As an example, to go into more numerical detail, \citet{kremer2020} predict 100 to 200 WD–WDs per $10^6$ M$_{\odot}$ in typical GCs, depending on the cluster's density, using \textsc{cosmic} \citep{cosmic}, a rapid population synthesis code derived from \textsc{bse}, and using \textsc{cmc} to model the effects of dynamical interactions. By comparison, \citet{thiele2023}, also using \textsc{cosmic} but looking at MW disk populations without dynamical interactions, predict on the order of 500 to 10,000 WD–WDs per $10^6$ M$_{\odot}$, depending on metallicity and binary evolution assumptions. Meanwhile, \citet{shara2006}, using \textsc{bse} and N-body dynamics, predict 500 WD–WDs per $10^6$ M$_{\odot}$ for a field population but 3000 WD–WDs per $10^6$ M$_{\odot}$ for a GC.

In general, both EM observations and simulation-based predictions in the literature seem inconclusive about the effects of dynamical interactions upon WD–WD binaries, including whether this would result in more or fewer WD–WDs occurring in GCs than in the field. Our own results broadly align with this in the sense that the discrepancy between our WD–WD predictions in this section and the \textsc{cmc}-\textsc{bse} predictions of \citet{kremer2018_gc_lisa} is of a similar magnitude (a factor of a few $\times$ 10) to the discrepancy between our results for the Magellanic Clouds and the \textsc{SeBa} and \textsc{bse} studies discussed in Sect. \ref{magellanic_discussion} – despite dynamical interactions being a factor for GCs but negligible for the Magellanic Clouds.

The consistency of the disparity across both the GC and Magellanic simulations suggests that the effect of dynamical interactions upon the total number of WD–WD binaries that would be detectable by LISA is likely smaller in magnitude than the effect of the differences in the treatment of stellar evolution between \textsc{bpass} and \textsc{SeBa}/\textsc{bse}. The reason for this is that, if the effect of dynamical interactions was in fact significant, we would expect the discrepancy between our results and previous research to be larger for the GCs than for the Magellanic Clouds. However, this is a very simple estimation and fully understanding the impact of the inclusion of dynamical interactions upon the binary population synthesis results merits further research.

\subsubsection{Ejections and cluster mass loss}

Aside from their effect upon the evolution of individual binaries, the lack of dynamical interactions in our simulations could also be affecting our results through their effect upon the overall mass of the cluster. Specifically, as described in Sect. \ref{initial_mass_calc}, we compute the initial mass of a cluster from the current mass using \textsc{bpass} data files that account for the mass lost by individual stellar systems through their evolution. However, there is an additional way in which GCs can lose mass, in that dynamical interactions can cause stars or compact remnants, particularly heavier objects like BHs, to be ejected from the cluster \citep{gcbhejection2}. A broad overview of the mechanisms whereby objects can be ejected from GCs is given in Sect. 2 of \citet{weatherford2023}.

Taking this effect into account would increase the mass lost by a GC over time, and thus for a given current mass would increase the computed initial mass, which would increase the present-day amount of binaries in our simulations. This could therefore account for some of the difference between our predicted numbers of binaries and those of the \textsc{cmc} simulations of \citet{kremer2018_gc_lisa}.

However, the magnitude of the mass loss due to dynamical ejections varies significantly between different simulations in the literature as it depends on numerous properties of the cluster involved, such as its mass, radius and Galactic location. One particular relevant parameter is the dynamical relaxation timescale, which is proportional to the cluster mass if its density is held constant. For example, \citet{shara2006}, from an N-body simulation of a single GC, found about $75\%$ of objects were ejected after 15 Gyr and $90\%$ after 20 Gyr, and the total mass of the cluster was only about $20\%$ of its initial mass after 16 Gyr. However, \citet{weatherford2023} performed \textsc{cmc} simulations of GCs, where the clusters most analogous to those of \citet{shara2006} – with similar densities and Galatocentric distances, but with eight times as many objects, leading to a slower dynamical timescale – had only 10-20$\%$ of objects ejected after 14 Gyr, with the cluster masses being about $50\%$ of their initial masses at that time.

We also note that \citet{pijloo2015} created a code to calculate the initial mass of a cluster based upon its current parameters using a combination of Monte Carlo simulations and semi-analytical methods; this code uses a simple approximation of mass loss from stellar evolution which is outlined in \citet{alexander2014}, but it could potentially be adapted to use the more detailed \textsc{bpass} mass loss calculations in the future.

\section{Conclusions} \label{conclusion_chapter}

In summary, we draw the following conclusions:
\begin{enumerate}
    \item The Magellanic Clouds and GCs are potential hosts of GW sources that would be detectable by LISA. The \textsc{bpass} simulations suggest the Magellanic sources would be a mixture of NS–WDs, WD–WDs and potentially other types of binaries such as BH–WDs, while the GC sources would be predominantly WD–WDs. This indicates that non-WD–WD binaries should not be ignored in GW analyses of the Magellanic Clouds.
    \item We find a few tens of times fewer detectable WD–WD binaries in the LMC than a comparable study by \citet{lmclisa} which used the stellar evolution/population synthesis code \textsc{SeBa}. A similar discrepancy was found by \citet{bpassmilkyway} when they compared their \textsc{bpass} model for the WD–WD binaries in the MW disk to the results of \citet{lamberts2018,lamberts2019}, who used \textsc{bse}, which indicates that the discrepancy originates from the stellar evolution codes.
    \item More specifically, we believe that the discrepancy between our results on the one hand and the earlier research which used \textsc{SeBa} and \textsc{bse} on the other originates from differences in the treatment of stellar evolution, particularly relating to CEEs and the stability of mass transfer. Some tests involving simulating a stronger CEE in \textsc{bpass} give us further confidence in this explanation.
    \item For GCs, our predictions using solely the isolated binary evolution of \textsc{bpass} give an expected number of detectable binaries of less than one even for the most massive GC, $\omega$ Centauri. These numbers are also a few tens of times lower than those predicted by \citet{kremer2018_gc_lisa}, who used \textsc{cmc} to simulate GC dynamics and \textsc{sse}/\textsc{bse} for stellar evolution. The discrepancy is larger for binary types other than WD–WDs, which are more likely to be affected by GC dynamical interactions.
    \item Different studies in the literature disagree on whether GC dynamical interactions increase or decrease the number of detectable WD–WDs, but the size of these effects is likely to be on the order of a few. Based on the consistency in the size of the discrepancy between \textsc{bpass} and other codes in detectable WD–WD numbers across both the Magellanic and GC simulations in our study, we also do not see a clear effect from the inclusion or exclusion of GC dynamics on WD–WDs. However, we do see that the stellar evolution differences affect the WD–WD rates in GCs by a factor of a few tens.
\end{enumerate}

\section*{Acknowledgements}

WGJvZ acknowledges support from Radboud University, the European Research Council (ERC) under the European Union’s Horizon 2020 research and innovation programme (grant agreement No.~725246), Dutch Research Council grant 639.043.514 and the University of Auckland. JJE acknowledges support from Marsden Fund Council grant MFP-UOA2131 managed through the Royal Society of New Zealand Te Apārangi.

\section*{Data availability}

The \textsc{bpass} data set can be downloaded from \url{https://bpass.auckland.ac.nz}. The \textsc{legwork} code can be found at \url{https://github.com/TeamLEGWORK/LEGWORK}. Other data underlying this article will be shared upon reasonable request to the corresponding author.

\bibliographystyle{aa}
\bibliography{references}

\end{document}